\documentclass[preprint,aps,amsmath,amssymb,superscriptaddress,footninbib]{revtex4}

\usepackage{graphicx}

\newcommand{\ev}[1]{\left\langle #1 \right\rangle}

\begin{document}

\title{Black hole entropy from the quantum atmosphere of bound gravitational fluctuations}

\author{Seth Major, Daniel Rodriguez, Thomas Takis}
\affiliation{Department of Physics, Hamilton College, Clinton NY
13323 USA}

\date{January 2025}

\begin{abstract}
Black hole entropy is identified with the counting of the dynamical degrees of freedom of trapped gravitational modes continually sourced by the Hawking-Unruh process. In the context of linear perturbations of Schwarzschild spacetime the density of states is derived from the orthogonality of states in the solution space of the Regge-Wheeler-Zerilli equation. The otherwise divergent energy and entropy is cutoff by the Planck scale closest approach of constantly accelerating observers near the horizon.  The thermal distribution of the trapped modes, which represent shape fluctuations in the near horizon geometry, store a significant fraction of the spacetime mass as observed from far away.  Unlike quasi-normal modes the modes are not directly observable outside of $\sim 3 M$ but, being external to the horizon, they affect the propagation of null rays near the black hole.  The characteristic frequencies, around 100 Hz for solar mass black holes, are discussed in relation to possible observations. 

\end{abstract}

\maketitle

\section{Introduction}

Do the degrees of freedom of a black hole leave an observational signature? An extraordinary array of physical systems have characteristic modes of oscillation such as the late time sound of the ringing of a bell.  The frequencies carry information on the geometry or shape of the bell. 
Similarly the ``shape" of atoms and molecules may be explored with spectroscopy. While black holes are intrinsically dissipative they have a rich structure of gravitational quasi-normal modes (QNM) \cite{qnmreview09,25_qnm_rev} which have now been observed in the gravitational wave observations of the black hole binary system GW250114 \cite{area_qnm}.

Since Bekenstein and Hawking showed that black holes have a large thermodynamic entropy proportional to the horizon area \cite{bekenstein1,bekenstein2,hawking} 
\begin{equation}
	S = \frac{A}{4 \, \ell_p^2}
\end{equation}
the dynamical degrees of freedom responsible for the entropy have been somewhat mysterious.  In this paper we present a calculation showing that the entropy of bound gravitational modes, trapped between the horizon and the curvature potential, scale correctly with area. 
If indeed these trapped modes fully account for the entropy of black holes, we show that the atmosphere of these modes represent a significant mass fraction of the spacetime as observed from far away. While the trapped modes are not directly observable far away, being exterior to the black hole horizon, the modes cause the near horizon geometry to fluctuate affecting null geodesics. These fluctuations open a window to observations the dynamical degrees of freedom that ``count" the states of black holes.

For the setting of the calculation we consider a family of constantly accelerating observers (CAOs) who maintain a uniform acceleration near the horizon (or a family of fiducial observers FIDOs \cite{zurek_thorne}). In Schwarzschild spacetimes these are just static observers at constant radius. The CAOs enjoy a proper acceleration of $- g \hat{e}^r$, $g = M/r^2 (1 - 2M/r)^{-1/2}$.  Due to the Hawking-Unruh effect near horizon CAOs are bathed in a thermal atmosphere at temperature  $\hbar g / 2 \pi$. (We use units where $G=c=k=1$.)  In the framework we also assume a fundamental discrete geometry suggested by the condensed area state in \cite{MRT} that yields a fundamental area $a_1 \sim \ell_P^2$. Since the acceleration $g$ diverges at the horizon the observers cannot reach the horizon.  We assume that the observers have an in-principle closest approach at the discreteness scale of the geometry and impose this cutoff on the modes. 

The calculation rests on the considerable insights gained during the long history  of ``quantum atmosphere" calculations for black holes, see for instance \cite{york,box,zurek_thorne,frolov_novikov,TR,oshita_afshordi}. We follow Frolov and Novikov in placing the dynamical degrees of freedom of black holes in physical fields - here we suggest it is entirely gravitational modes  - that cannot be observed far away \cite{frolov_novikov}. Counting these degrees of freedom gives the Bekenstein-Hawking entropy.  In \cite{york} York suggested that quasi-normal modes were the source of the dynamical degrees of freedom of a black hole. To ensure that counting these modes gave the Bekenstein entropy he required a sum over the vanishing modes over the lifetime of the black hole. Here we show that the dynamical degrees of freedom are at a given moment of time.  't Hooft showed that a simple ``brick wall" model using Dirichlet boundary conditions gave the correct scaling of entropy with area \cite{box}. We recover the form of the density of states of the brick wall quantization in the high frequency limit.  In this framework we compute the {\em gravitational} thermal atmosphere while the other degrees of freedom of matter and radiation sectors contribute to the regular thermal entropy of the matter sector. 

As for the underlying constituents of geometry we use the kinematic results of loop quantum gravity, particularly the foundational nature of the area spectrum \cite{LCgeom,AbhayJurekArea}. The entropy and energy diverge without a cutoff. A state of condensed geometry derived in \cite{MRT} provides a framework for a geometric cutoff.  However, for this work only the existence of a fundamental area $a_1$ is important. We set $a_1 = \alpha \hbar$.  

  

Aside from the assumption of the granularity of geometry this work rests on additional approximations.  We use linear perturbation theory around Schwarzschild spacetime, take the near horizon limit to obtain a simple form of the equations of motion, and rely on high and low frequency limits for analytic control. For these reasons we regard this proposal as a ``proof of principle" calculation rather than a definitive treatment of the spectrum, energy and entropy of these trapped gravitational modes. We will nevertheless show that the trapped modes (1) can account for the Bekenstein-Hawking entropy, (2) store a significant fraction of the mass of the black hole spacetime, and (3) have a discrete spectrum that is potentially observable.


In the next section we describe the Schr\"odinger-like equation that governs the trapped modes in the linear approximation. Requiring that the solutions vanish far away and that neighboring solutions be orthogonal gives a quantization condition and the density of states. In section \ref{fluctuations} we derive the entropy and energy of the trapped modes showing that they can account for the Bekenstein-Hawking entropy in the low and high frequency limits.  We also derive the dominant frequencies for the trapped modes in the high and low frequency limits.  Due to the temperature scale, the low lying modes are favored and we use these to explore possible observations in section \ref{observations}. 

\section{trapped Gravitational Modes}
\label{trapped}

\begin{figure}
\label{potential}
	\includegraphics[scale=.7,angle=0]{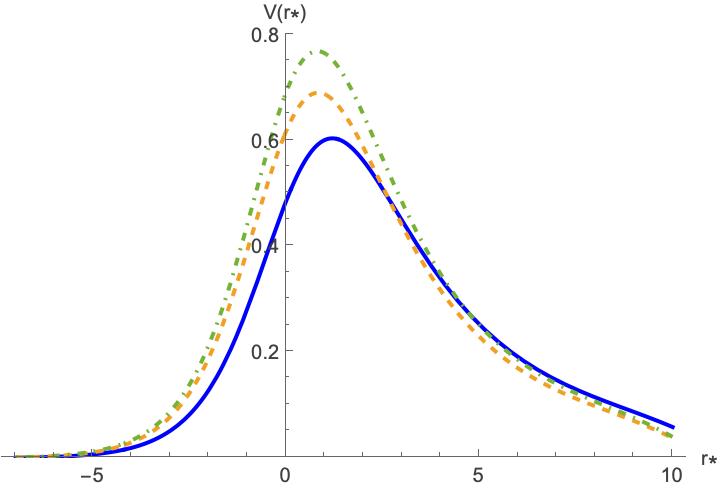}
\caption{ \label{Vfig} Plots of the potential $V_{s=2}(r_*)$ for $\ell=2$ (solid), $V_{s=2}(r_*)/15$ for $\ell=8$ (dashed), and $V_{s=2}(r_*)/800$ for $\ell=64$ (dot-dashed). The potentials have been rescaled to ease the comparison between different angular momenta showing that for large angular momentum the potential peak rises as $\ell^2$ and shifts inward compared to $\ell=2$ potential. The Schwarzschild radius $r_S = 2M$ has been set to 1. The horizon and the CAOs are on the left; asymptopia at $r\rightarrow \infty$ is to the right.}
\end{figure}

To investigate the fluctuation in the shape of the near-horizon geometry we start with linear perturbations $h_{\mu \nu}$ around a background Schwarzschild geometry with metric $g_{\mu \nu}$. After a decomposition into tensorial spherical harmonics, the Einstein equations for the linear perturbations take the form  $\Box_g h_{\mu \nu} =0$. In the Regge-Wheeler gauge, the perturbations separate into even and odd parity solutions (see \cite{qnmreview09} for a review).  The metric components in this gauge may be all derived from the two radial wavefunctions $\Psi^\pm_{s=2}$ which satisfy the Regge-Wheeler-Zerilli (RWZ) equation
\begin{equation}
	\label{RW}
	\frac{d^2 \Psi^\pm_{s=2}}{dr_*^2} + \left( \omega^2 - V_s\left[ r(r_*) \right] \right) \Psi^\pm_{s=2}(r_*) = 0,
\end{equation}
where the potential is
\begin{equation}
	\label{potentialV}
	V_s(r) = \left( 1 - \frac{2M}{r} \right) \left[ \frac{\ell(\ell+1)}{r^2} + (1-s^2) \frac{2 M}{r^3} \right].
\end{equation}
The RWZ equation is in terms of the radial tortoise coordinate 
\[
	r_* = r + 2M \ln (r/2M -1) \text{ and } \frac{d r}{dr_*} = 1 - \frac{2M}{r}.
\]
The potential $V_s\left[ r(r_*) \right] $ is plotted in figure \ref{Vfig} for three angular modes. The Schr\"odinger-like RWZ equation (\ref{RW}) governs scalar ($s = 0$), electromagnetic ($s = \pm 1$) and Regge-Wheeler or gravitational ($s = 2$) radial perturbations \cite{qnmreview09}. The signs designate the axial (partity odd) Regge-Wheeler perturbations and to the polar (partity even) Zerilli perturbations. Remarkably the Regge-Wheeler and Zerilli modes are isospectral in the Schwarzschild background \footnote{Using the boundary conditions $\Psi^{\pm} \sim e^{-i \omega r_*}$ as $r \rightarrow r_S$ and $\Psi^{\pm} \rightarrow 0$ as $r \rightarrow \infty$ in the argument in Appendix A of Ref. \cite{qnmreview09} one may show that the two trapped Regge-Wheeler and Zerilli perturbations isospectral.  The more subtile case of algebraically special modes is left to further work.}.

An algebraically easier route to the Regge-Wheeler-Zerilli equation is through scalar perturbations $\Psi$ with equation of motion   
\[
	\Box_g \, \Psi = 0 \text{ or } - \frac{1}{\sqrt{-g}} \partial_\mu \left( \sqrt{-g} g^{\mu \nu} \partial_\nu \Psi \right) = 0 .
\]
Assuming a solution of the form
\begin{equation}
	\label{modebasis}
	\Psi (t,r,\theta,\varphi)= \sum_{n \ell m} c_{n \ell m} e^{-i \omega_n t} \frac{u_{n \ell}(r) }{r} Y_{\ell m}(\theta, \varphi) 
\end{equation}
gives the RWZ equation (\ref{RW}) for $u_{n \ell}$ in the tortoise coordinate $r_*$ with the potential $V_{s=0}(r)$.

There is a surprising amount of physics in the potential of (\ref{potentialV}). It contains the complex structure of the QNMs, defined as infalling at the horizon and outgoing at infinity.  Here we consider trapped gravitational modes that are sourced by the Hawking-Unruh process and observed by the CAOs.  The inward traveling modes fall directly into the black hole while outward traveling modes reflect off the potential (\ref{potentialV}) and fall back into the black hole. The modes are infalling at the horizon and vanishing at infinity. Given the similar physics and the intricate nature of the QNM overtones we expect that the trapped modes have complex frequencies but here we focus on their real part.

The inner barrier of the potential rises exponentially for these trapped modes.  For simplicity we study the modes in the near horizon limit where  
\begin{equation}
	\label{approx_potentialV}
	V_s(r)  \simeq \frac{\ell(\ell+1)+ 1-s^2}{4 M^2} \exp(r_*/2M) = \left( \frac{\lambda}{2 M} \right)^2 e^{r_*/2M}.
\end{equation}
We define $\lambda^2 = \ell(\ell+1) +1 -s^2$. With this parameterization the differential equations for the gravitational scalars $\Psi^\pm_{s=2}$, equation  (\ref{RW}), and for the scalar modes $u_{n \ell}$ are identical up to $s$.  For the remainder of the paper we denote the gravitational scalars by $\Psi(t,r,\theta,\varphi)$ and set $s=2$. The factor $\lambda/2M$ is an effective transverse wave number $k_\perp$. Inside the potential well excitations near the horizon satisfy
\begin{equation}
	\label{insidecondition}
	\frac{\omega^2}{1 - 2M/r} \geq  \frac{\lambda^2}{4 M^2},
\end{equation}
that is, the local mode frequency as measured by CAOs must be greater of equal to the transverse wave number, $\omega_{loc} \geq k_\perp$, classically.  This condition plays a key role in the density of states.

In the horizon's asymptopia, when $r_* \rightarrow - \infty$, the mode equation simplifies to simple harmonic form and the solutions are in- and out-moving waves in flat space. At large angular momenta, where the bulk of the states lie, $\lambda \simeq \ell$ and the spectra of trapped scalar and gravitational modes are equivalent.  




To solve the RWZ equation we let
\begin{equation}
	\label{z}
	z = 2 \lambda \exp (r_*/4M)
\end{equation}
and $\Omega = 4 M \omega$. Denoting the $z$ derivative with a prime, the RWZ equation (\ref{RW}) with the near horizon approximate potential (\ref{approx_potentialV})  becomes the near horizon mode equation 
\begin{equation}
	\label{mode_equation}
	z^2 u''_{n \ell} + z u'_{n \ell} + \left( \Omega^2 -z^2 \right) u_{n \ell} = 0.
\end{equation}
This is the modified Bessel equation of imaginary order. (We study the solutions for real $\Omega$.) The coordinate $z$ vanishes at the horizon and diverges as $r \rightarrow \infty$. The solutions are $u_{n \ell} (z) = A \, I_{i \Omega} (z) + B \, K_{i \Omega} (z)$. The $ I_{i \Omega} (z)$ solutions diverge at large $z$ so only the $K_{i \Omega}$ are physical.   Although at the horizon the modes are infalling, the high acceleration observers outside the horizon are bathed in a thermal bath of modes at temperature $\hbar g/2 \pi$. The $u \sim K_{i \Omega}$ solutions means that at the observers' position the radial current, $j^r = i ( u^* \partial^r u - u \partial^r u^*)$, identically vanishes, as it must for stationarity.

We parameterize the near horizon limit with 
\begin{equation}
	r_{min} = 2M \left( 1+\frac{\epsilon^2}{4} \right) \text{ and } z_{min} \simeq \sqrt{e} \lambda \epsilon.
\end{equation}
(As we will see below the maximum value of $z_{min}$ is $\sqrt{8 \pi e} $.)

To model the thermal atmosphere of trapped gravitational modes we need to characterize them and derive a density of states. In the near horizon and large angular limit we can obtain this characterization from the orthogonality of neighboring states in the Klein-Gordon inner product for solutions of the mode equation (\ref{mode_equation}). (See \cite{TR} for the analysis in freely falling frames.). For a spatial slice for constant Schwarzschild coordinate $t$ and its normal $n^\alpha$ (the 4-velocity of CAOs)
\begin{equation}
	\ev{ \Psi_1, \, \Psi_2 } = i \int_{\Sigma_t}  \, n^\alpha \left( \Psi_1^* \partial_\alpha \Psi_2 - \Psi_2 \partial_\alpha \Psi_1^* \right) \sqrt{h}  \, d^3x .
\end{equation}
As in equation (\ref{modebasis}) we expand in spherical harmonics and integrate over the sphere obtaining
\begin{equation}
	\ev{ \Psi_1, \, \Psi_2 } =  \left( \omega_1 + \omega_2 \right)  e^{i (\omega_1 - \omega_2) t} \int_{r_{*min}}^\infty d r_*  \,  u_{\omega_1 \ell}^* u_{\omega_2 \ell} .
\end{equation}
Changing variables to $z$ and using the modified Bessel solutions this becomes \footnote{We note that the inner product with vanishing $z_{min}$ has the form 
\[
	\int_0^\infty \frac{dz}{z} K_{i \nu} (z) K_{i \nu'} (z) = \frac{\pi^2}{2 \nu \sinh(\pi \nu)} \delta(\nu - \nu')
\]
for pure imaginary orders so the inner product requires regulation.}  
\begin{equation}
	\begin{split}
	\ev{ \Psi_1, \, \Psi_2 } &= \left( \Omega_1 + \Omega_2 \right) e^{i (\omega_1 - \omega_2) t}  \int_{z_{min}}^\infty \frac{dz}{z} K^*_{i \Omega_1} (z) K_{i \Omega_2} (z) \\
	&=  \left( \frac{e^{i (\omega_1 - \omega_2) t}  }{\Omega_1 - \Omega_2} \right)  \left[ z \left( K'_{i \Omega_1}  K_{i \Omega_2}   - K'_{i \Omega_2}  K_{i \Omega_1} \right) \right]_{z_{min}} .
	\end{split}
\end{equation}
In the near horizon region when $z \rightarrow 0+$ the modified Bessel functions have the asymptotic form \cite{dunster}
\begin{equation}
	K_{i \Omega}(z) = \sqrt{ \frac{\pi}{\Omega \sinh(\pi \Omega)}} \sin \left[ \Omega \ln (z/2) - \phi_\Omega \right] + {\mathcal O}(z^2),
\end{equation} 
where $\phi_\Omega = \arg \left( \Gamma(1+i \Omega) \right)$. Using the leading order form of the approximation in the inner product gives
\begin{equation}
	\begin{split}
	\ev{ \Psi_1, \, \Psi_2 }  &\propto \sin \left[ \left( \Omega_1 + \Omega_2 \right) \ln ( z_{min}/2 ) -\phi_{\Omega_1} -  \phi_{\Omega_2} \right] \\ &- \left(
	\frac{\omega_1 + \omega_2}{\omega_1 - \omega_2} \right) \sin \left[ \left( \Omega_1 - \Omega_2 \right) \ln ( z_{min}/2 ) -\phi_{\Omega_1} +  \phi_{\Omega_2} \right].
	\end{split}
\end{equation} 
Orthogonality of states for $\Omega_1$ and $\Omega_2$ requires both sines to vanish. The two conditions determine the mode frequency quantization
\begin{equation}
	\label{orthog}
	2 \Omega \ln (z_{min}/2) - 2 \phi_\Omega = n \pi
\end{equation}
for non-vanishing integer $n$. 

Some insight can be gained by exploring low and high order limits of $\phi_\Omega$. For large $\Omega$, $\phi_\Omega \sim \Omega \ln \Omega$ so that (\ref{orthog}) gives
\begin{equation}
	\label{spectrumlargeorder}
	 n \pi = 8 M \omega_{n \ell}  \ln \left( \frac{8 M \omega_{n \ell}}{z_{min}(\ell)} \right),
\end{equation} 
the quantization of the trapped modes in this limit.  Up to a factor of two, this is the same quantization condition of the brick wall model \cite{box} when one imposes the Dirichlet boundary conditions $K_{i \Omega}(z_{min}) = 0$ in this limit. Thus we recover the boundary conditions of `t Hooft's  ``elementary exercise" in \cite{box}; the brick wall modes are physical in this limit \footnote{This may be further verified by using a WKB argument in the Appendix.}.  

The minimum value of $z$ is found from the scale of discrete geometry. The proper distance from the horizon to the coordinate $r_{min}$ is approximately
\[
	d_{phys} = \int_{2M}^{r_{min}} \sqrt{g_{rr}} dr \simeq 2 M \epsilon .
\]
The area of the surface of the family of CAOs at this radius is $4 \pi r_{min}^2 \simeq 4 \pi  d_{phys}^2 /\epsilon^2$.  We set the area $d_{phys}^2$ to the area discreteness scale $a_1$.  Thus, 
\begin{equation}
	\label{epsilonarea}
		\epsilon^2 = \frac{4 \pi a_1}{A} \simeq \alpha \cdot 3.5 \times 10^{-65} \left( \frac{M_\odot}{M} \right)^2
\end{equation}
and $z_{min} =  \lambda \sqrt{e} \, \epsilon$. These quantities are extraordinarily small unless the angular momentum is large in which case $z_{min}$ may be ${\mathcal O}(1)$.

From this value for $z_{min}$ and the quantization at large order (\ref{spectrumlargeorder}) the density of states is thus 
\begin{equation}
	\label{highg}
	g(\omega, \ell) = \frac{\partial n}{\partial \omega} = \frac{8 M}{\pi} \ln \left( \frac{8 \sqrt{e} \, M \omega}{\lambda \epsilon} \right).
\end{equation}
In the same limit the zeros of the modified Bessel functions also give this density of states (up to a factor of 2) so the orthogonality property reproduces the Dirichlet condition. Since reflections of any constant phase produce nodal surfaces it is perhaps not surprising that orthogonality recovers the brick wall boundary condition.
 
In the limit of small $\Omega$, $\phi_\Omega \sim - \gamma_o \Omega$, where $\gamma_o$ is the Euler-Mascheroni constant, so the orthogonality condition (\ref{orthog}) gives
\begin{equation}
	\label{lowomega}
	\omega_n = \frac{n \pi}{8 M} \frac{1}{ \ln \left( \frac{2 \sqrt{e}}{ e^{\gamma_o} \lambda \epsilon} \right)} 
\end{equation}
giving the density of states in the low order limit 
\begin{equation}
	\label{lowg}
	\tilde{g}(\ell) = \frac{8 M}{\pi} \ln \left( \frac{2 \sqrt{e}}{ e^{\gamma_o} \lambda \epsilon} \right),
\end{equation}
which no longer scales with the frequency. With the density of states in both limits we can now compute the thermodynamic quantities.

\section{Fluctuations in shape}
\label{fluctuations}
 
We explore the thermodynamics of near horizon shape fluctuations using usual statistical techniques. Given the approximations we have employed (linear perturbations, near horizon geometry, high and low order)  we do not expect this to give a precise account of the Bekenstein-Hawking entropy.  Rather we wish to show that the $S \propto A$ scaling is correct. The partition function for the liner perturbations is
\begin{equation} 
	Z = \prod_{n,\ell} \left( 1 - e^{- \beta \hbar \omega_{n \ell} } \right)^{-(2\ell +1)}.
\end{equation}
The energy and entropy of the quantum atmosphere of geometry phonons takes the usual forms 
\begin{equation}
	\begin{split}
		U &= \sum_{n,\ell} (2 \ell +1) \frac{\hbar \omega_{n \ell} }{ e^{\beta \hbar \omega_{n \ell}} -1 } \\
		S &=  \sum_{n,\ell} (2 \ell +1) s(\omega_{n \ell}) \text{ with }
		s(\omega) = \left[ \frac{\beta \hbar \omega}{e^{\beta \hbar \omega} - 1} - \ln \left( 1 - e^{-\beta \hbar \omega} \right) \right].
	\end{split}
\end{equation}
Perhaps the conceptually clearest frame for this calculation is that of the high acceleration observer for whom the modes are blueshifted and of the form $\omega \sim g \nu$ for dimensionless frequency $\nu$. These observers are bathed in thermal atmosphere at the Hawking-Unruh temperature $g \hbar / 2 \pi$.  Because the Boltzmann weight depends on the dimensionless energy, $\beta \hbar \omega = 2 \pi \nu$.  The distribution is invariant under changes in the CAOs acceleration $g$.  In the near horizon limit the results are universal \footnote{This is well known but it remains to be seen if a local definition of energy and entropy hold more generally. See \cite{Eg} for suggestions that local definitions may exist.}. Since we wish to compare the trapped mode frequencies as measured far away, we will work with frequencies and temperatures appropriate for the asytompic form, using asymptotic frequencies $\omega \sim 1/M$ and Hawking temperature $\hbar/ 8 \pi M$ for the distributions. This temperature is the ``would be" Hawking temperature of the thermal atmosphere  were one able to observe the trapped modes far away.

We will compute the thermodynamic quantities in both high and low order limits.  The high order limit recovers the brick wall results and the low order is more directly relevant to the trapped modes. Starting with the high order limit the entropy may be approximated with integrals
\begin{equation}
	S \simeq \int_0^{\omega_{max}} d \omega \int_2^{\ell_{max}} (2 \ell +1) g(\omega, \ell) s(\omega).
\end{equation}
The maximum angular momentum $\ell_{max}$ is set by two conditions. First, to ensure that the modes are trapped inside the potential (\ref{insidecondition}) then $\ell_{max} (\omega) =  4 M \omega / \sqrt{e} \, \epsilon$ at leading order \footnote{This follows once one recalls that $(1 - 2M/r) \simeq \epsilon^2/4$.}. In the high order limit this condition also ensures that the density of states remains positive. Second following a Debye argument, the absolute number of angular modes is bounded from above by the total number $N$ of geometric tiles of the fundamental area $a_1$ (set by the underlying geometric condensate in \cite{MRT}).  Thus,
\[
	\sum_{\ell=2}^{\ell_{max}} (2 \ell + 1) d \ell \simeq \int_{2}^{\ell_{max}} (2 \ell +1) = 2 N.
\]
But since the area is $A = a_1 N$,
\begin{equation}
	\ell^2_{max} = \frac{2 A}{a_1}.
\end{equation}

Using the high order density of states from (\ref{highg}), $\lambda \simeq \ell$, and $\ell_{max}  =  4 M \omega / \sqrt{e} \, \epsilon$ the angular integration becomes
\begin{equation}
	\frac{8 M}{\pi} \int_2^{\ell_{max}} (2 \ell +1) \ln \left( \frac{8 \sqrt{e} \, M \omega}{\lambda \epsilon} \right) d \ell  =  \frac{64 M^3 \omega^2}{\pi \epsilon^2} \left[ 1 + {\mathcal O}(1/\ell_{max} ) \right].
\end{equation}
This is equivalent to integrating over the transverse momenta. For large black holes the integrand peaks at $\omega A /\sqrt{a_1}$.  Dropping the sub-dominant term, the entropy becomes
\begin{equation}
	\begin{split}
	S &\simeq \frac{64 M^3}{\pi \epsilon^2} \int_0^{\omega_{max}} d \omega \, \omega^2 s(\omega) \\
	&\simeq \frac{64 M^3}{\pi (\beta \hbar)^3 \epsilon^2} \int_0^\infty dx \, x^2 \left[ \frac{x}{e^x-1} - \ln \left( 1 - e^{-x} \right) \right].
	\end{split}
\end{equation} 
As discussed above the temperature is the would-be Hawking temperature of the modes $\hbar/8 \pi M$, if they were observable far away.  Due to the entropy density, which is peaked around $\omega \sim 0.1/M$ we can extend the upper limit $\omega_{max}$ to infinity without significantly affecting the result.  The integral evaluates to $4 \pi^4/45$. So   
\begin{equation}
	\label{entropy}
	S \simeq \frac{1}{90 \epsilon^2} = \left( \frac{1}{90 \pi \alpha} \right) \left(  \frac{ A}{4 \, \ell_P^2} \right),
\end{equation} 
using the near horizon cutoff in (\ref{epsilonarea}). The entropy scales with area and so the trapped modes can account for the Bekenstein-Hawking entropy.  Matching this result requires $\alpha = 1/90 \pi$ when it is expected that $\alpha \sim {\mathcal O}(1)$. Likely this undercounting is due the above approximations, but since we are applying a cutoff at the Planck scale renormalization of the gravitational constant could also play a role.

The calculation of the energy is similar. With the same density of states and approximations 
\begin{equation}
	\label{energydist}
	U \simeq \frac{64 M^3}{\pi \hbar^3 \beta^4 \epsilon^2} \int_0^\infty  \frac{x^3}{e^x-1} dx = \frac{\hbar}{240 a_1} M 
\end{equation}
(The integral evaluates to $\pi^4/15$.). Matching the Bekenstein-Hawking form of the entropy the energy stored in the trapped modes is
\begin{equation}
	\label{renomrenergyhigh}
	U = \frac{3}{8} M,
\end{equation}
as first noticed in \cite{box}. 

The calculation for lower order limit when $\Omega <1$ proceeds in the same manner.  Using the density of states of (\ref{lowg}) in the angular integration gives
\begin{equation}
	\begin{split}
	\frac{8 M}{\pi} \int_2^{l_{max}} (2 \ell +1) \tilde{g}(\ell) d\ell &=  
	\frac{8 M}{\pi} \int_2^{\ell_{max}} \ln \left( \frac{2}{ e^{\gamma_o -1/2} \lambda \epsilon} \right) d \lambda \\
	\simeq \frac{128 M^3}{a_1}. 
	\end{split} 
\end{equation}
Here we have used $\ell_{max} = \sqrt{e A / \pi e^{2 \gamma_o} a_1}$ to ensure the positivity of the density of states.
The angular integration no longer depends on the frequency. The entropy is then
\begin{equation}
	S = \frac{128 M^3}{ (\beta \hbar) a_1} \int_0^\infty dx \, \left[ \frac{x}{e^x-1} - \ln \left( 1 - e^{-x} \right) \right] = \left( \frac{4}{3 \alpha} \right) \left(  \frac{ A}{4 \, \ell_P^2} \right).
\end{equation}
Although the approximation overestimates the entropy, the low order limit accounts for the $S \propto A$ scaling of the Bekenstein-Hawking entropy.  The energy of the low order distribution is   
 \begin{equation}
 	\label{energydistlow}
 	U = \frac{128 M^3}{ (\beta^2 \hbar) a_1} \int_0^\infty dx \, \frac{x}{e^x-1} = \frac{M}{3 \alpha}.
\end{equation}
If the entropy accounts for the total then $\alpha = 4/3$ and the energy stored in the fluctuations is $M/4$. The dimensionless energy density is also monotonically decreasing and so favors the lowest modes, as one would expect from simple counting arguments.
 
Bound and stationary, the energy of these modes is associated to the compact source for distant observers.  Outside of $r \simeq 3M$ the Schwarzschild metric is sourced by a total mass of $\tilde{M} = 11/8 \, M$ in the high order case and $\tilde{M} = 5/4 \, M$ in the lower order case.  In the high order limit the energy expression (\ref{energydist}) shows that the gravitational radiation has a Planck spectrum. The peak frequency is
\begin{equation}
	\label{peakhighorder}
	\omega_{*} = 2.82 \cdot \frac{11}{8} \cdot \frac{1}{8 \pi \tilde{M}} \simeq 0.15 \frac{1}{\tilde{M}} 
\end{equation}
or a frequency about
\begin{equation}
	f_* = 5.0 \left( \frac{M_{\odot}}{ \tilde{M} } \right) \text{ kHz}
\end{equation}
for distant observers using the Schwarzschild metric for the full combined mass $\tilde{M}$.  

In the lower order limit the distribution of (\ref{energydistlow}) is a monotonically decreasing function so the fundamental $(n=1, \, \ell=2)$ frequency 
\begin{equation}
	\begin{split}
	\omega_{12} &=  \frac{\pi}{8  M} \frac{1}{ \ln \left( \frac{2 \sqrt{e}}{ \sqrt{3} e^{\gamma_o} \epsilon } \right)} 
	 \simeq  \left( \frac{5 \pi}{32 \tilde{M}} \right) \left[ \ln \left( \frac{ 8 \sqrt{e} \tilde{M}}{5 e^{\gamma_o} \sqrt{\hbar}} \right) \right]^{-1}  \\ 
	 &\simeq \left( 1.1 \text{ kHz} \right) \left( \frac{M_\odot}{\tilde{M}} \right) \left[1 + \frac{1}{87} \ln \left( \frac{\tilde{M}}{M_\odot} \right) \right]^{-1}
	 \end{split}
\end{equation}
dominates the sum \footnote{The maximum frequency, when  $\lambda \epsilon \sim \sqrt{8 \pi}$ is on the same order as the high order frequency (\ref{peakhighorder}).}. 
This corresponds to a frequency of 
\begin{equation}
	f_{12} \sim 180 \left( \frac{M_{\odot}}{\tilde{M}} \right) \text{ Hz}.
\end{equation}
In both cases the excited trapped modes are low lying modes inside the potential. The energy distribution and a simple counting argument - roughly the correct entropy can be obtained by assembling a configuration out of trapped frequencies of $(4 \pi M)^{-1}$ - suggest that the lower frequency dominates the trapped mode excitation. 

\section{Possible observational scenarios}
\label{observations}

As equations (\ref{renomrenergyhigh}) and (\ref{energydistlow}) show considerable energy is concentrated in the near horizon region between the horizon and the top of the potential barrier at $r \sim 3M$, if the trapped gravitational modes account for the gravitational entropy.  Due to the energy distributions the effect is universal; the product of frequency and mass is constant source to source. Astrophysically the best chance for observation of these modes are bright sources of light near the horizon.  Time series from such sources could reveal the fluctuations. But this is challenging. Most astrophysically relevant black holes rotate so the framework needs generalizing to the Kerr geometry.  Additionally, the source is compact; there are similar frequencies in the system including the orbital frequency at the photon ring; the probability of emission to infinity drops below 1/2 inside of $3 M$; and the astrophysical environment is modified due to infalling matter. 

There are several areas where one might be able to observe the characteristic trapped mode frequencies including quasi-periodic oscillations in X-ray binary systems, flare events around Sgr A*, movies of black hole shadows, gravitational wave observations,  and analog black hole systems.  Perhaps currently the most promising is observations of X-ray binaries. (See \cite{qpos_smith,qpos_ingram} for reviews.) In black hole - star binary systems when the star approaches the Roche limit, close enough for the tidal forces to pull mass from the star, the inflowing matter can reach the Eddington luminosity limit, when outward radiation pressure matches the gravitational attraction, and last for several months. 
Quasi-periodic oscillations (QPOs) have been observed in X-ray binaries.  It is thought that ``variations are likely to arise from quite near the black hole itself, and exhibit frequencies that scale inversely with the black hole mass" \cite{qpos_smith}. The (high frequency) QPOs have timescales of a few hundred Hz.  Promisingly, there is a correlation between the frequency of QPOs (both high and low frequencies) and the black hole mass spanning many orders of magnitude from stellar mass to supermassive black hole sources. Since these systems often have spins moderate to high spin, the framework should be extended to include spin to make detailed comparisons.  In the case of the supermassive black hole source Sgr A* the period would be a few hours. Signals with this time signature have been observed by ALMA \cite{alma}.


\section{Discussion}

We have explored trapped linear gravitational perturbations of Schwarzschild spacetimes.  In the near horizon limit the equation of motion of these perturbations may be solved. We study the solutions with real frequencies. Without regulation, the Klein-Gordon inner product of these trapped modes is divergent.  We assume the near horizon limit is cutoff by discrete geometry. Requiring orthogonality of neighboring modes results in a set of modes.  In the high frequency limit we recover the form of the density of states used in the brick wall model (or in a WKB approximation). In the low frequency limit we find a uniform density of states.  In both cases the resulting entropy scales with area.  If the black hole entropy is accounted for by these trapped fluctuations then they store a large fraction of the mass observed far away. The frequency of these oscillations is around 180 Hz for a solar mass black hole and the product of mass times frequency has a universal form. Such frequencies have been observed in QPOs in X-ray binary systems.  Scaled up to the mass of Sgr A* the fluctuations would have a period on the order of a few hours, such as has been observed by ALMA.    

The above calculation indicates that in an accelerating frame the gravitational entropy of back hole spacetimes may be accounted for by trapped gravitational modes sourced by the Hawking-Unruh effect.  This suggests that for highly accelerating CAOs near a black hole who are bathed in a thermal bath of geometric phonons, the shape of near horizon geometry fluctuates dramatically. Critically, the framework requires a cutoff which we take to be at the Planck scale. 

The results rest on a series of approximations including linear perturbations, the near horizon limit, approximating the potential $V_s$, and  low and high frequency limits. Further work to check this ``proof of principle" calculation includes computing the frequency $f_{12}$ numerically and employing methods to explore complex frequencies such as non-Hermitian methods, such as in developed by Zel'dovich in \cite{zeldovich}. This would address the mode spectrum for the trapped modes directly rather than using the orthogonality of states to obtain the density of states. 



We have computed the gravitational entropy from gravitational trapped modes, This appears to sidestep the species problem where  one finds the entropy depends on the number of allowed species of field.  For CAOs indeed additional fields would be excited, depending on the temperature and thus $g$, but in observing the bath of photons, electrons, positrons, etc. CAOs would count these degrees of freedom as regular radiation and matter degrees of freedom contributing to the matter rather than gravitational entropy. 

Outside of the stationary frame of the CAOs the story is less clear. Where are the gravitational modes produced? While ``obvious" for the high acceleration CAOs - they are bathed in a thermal bath of radiation at the Unruh temperature - it is less so for other frames.   We presume that in the freely falling frame the entropy can equivalently be derived by considering entanglement entropy \cite{sorkin_entanglement,entanglement2,TR}. As a quantum gravity effect one might hope that the area cutoff would be applied in all frames.  But \cite{TR} shows that in a freely falling frame the Planck scale cutoff does not yield the correct scaling of the entropy with area.

In \cite{bekenstein2} Bekenstein describes black hole entropy  ``...in the sense of inaccessibility of information about its internal configuration." It seems remarkable that a simple re-interpretation of Bekenstein's description - ``internal" is inside the potential barrier rather than beyond the horizon - could both explain the black hole entropy and offer a possible observational opportunity to view the entropic degrees of freedom. 

\begin{acknowledgments}
We thank Niayesh Afshordi, Ted Jacobson, Ryley McGovern, Jack Steiner, and Madhavan Varadarajan for helpful conversations and Hamilton College for generous financial support.
\end{acknowledgments}

\appendix 

\section{WKB Approximation}

Following \cite{box} we use a WKB approximation to compute the density of states. The radial wave number at large angular momentum satisfies
 \begin{equation}
 	k^2(r_*) = \omega^2 -\frac{\ell^2}{4 M^2} e^{r_*/2M} 
\end{equation}
in the near horizon limit. The Bohr-Sommerfeld-like quantization condition is
\begin{equation}
	n (\omega, \ell) \pi = \int_{r_{*min}}^{r_{*turn}} k(r_*) \, d r_* = \int_{r_{*min}}^{r_{*turn}} dx_* \left(\omega^2 -\frac{\ell^2}{4 M^2} e^{r_*/2M}\right)^{1/2}.
\end{equation}
The outer limit of the these trapped modes is the classical turning point of the effective potential (\ref{approx_potentialV}), $r_{*turn} = 4 M \ln \left( 2 M \omega/\ell \right)$, where the modes begin to exponentially decay.  (There is an exception which we do not model here, modes near the peak.  Since 1/2 of these modes leak out to infinity these quasi-normal modes have been well studied,  for instance in the beautiful WKB approximations of \cite{schutz_will}.)  The area scale $a_1$ provides the inner cutoff as in (\ref{epsilonarea}) giving $r_{*min} = 4 M \ln (2 \pi a_1 / A)$. With the change of variables $\xi = \ell e^{r_*/4M}/2 M \omega$ the integral is equivalent to 
\begin{equation}
	\begin{split}
		n \pi &= 4 M \omega \int_{\ell/4 M \omega \epsilon}^1 \frac{\sqrt{1 - \xi^2}}{\xi} d \xi  \\
		& = 4 M \omega \left[ \ln \left( \frac{8 M \omega}{e \ell \epsilon} \right) + {\mathcal O}(\epsilon^2) \right],
	\end{split}
\end{equation}
which in leading order gives the density of states 
\begin{equation}
	g_{WKB} (\omega, \ell) = \frac{ \partial n}{\partial \omega} = \frac{4 M \omega}{\pi} \ln \left( \frac{8 M \omega }{ e \ell \epsilon} \right) .
\end{equation}
This is essentially $g(\omega, \ell)/2$ of (\ref{highg}).

\end{document}